\documentclass[aps,prc,twocolumn,superscriptaddress,showpacs,amsmath]{revtex4}
\usepackage{graphicx}
\usepackage{bm}
\usepackage{multirow}

\begin{document}

\title{$\gamma$ strength functions in $\rm ^{60}Ni$ from two-step cascades following proton capture}

\author{A.~Voinov}
\email{voinov@ohio.edu} \affiliation{Department of Physics and Astronomy, Ohio University, Athens,
OH 45701, USA}

\author{S.M.~Grimes} \affiliation{Department of Physics and Astronomy, Ohio University, Athens,
OH 45701, USA}

\author{C.R.~Brune} \affiliation{Department of Physics and Astronomy, Ohio University, Athens,
OH 45701, USA}

\author{M.~Guttormsen}
\affiliation{Department of Physics, University of Oslo, N-0316 Oslo, Norway}

\author{A.C.~Larsen}
\affiliation{Department of Physics, University of Oslo, N-0316 Oslo, Norway}

\author{T.N.~Massey} \affiliation{Department of Physics and Astronomy, Ohio University, Athens,
OH 45701, USA}

\author{A.~Schiller} \affiliation{Department of Physics and Astronomy, Ohio University, Athens, OH 45701, USA}

\author{S.~Siem}
\affiliation{Department of Physics, University of Oslo, N-0316 Oslo, Norway}

\begin{abstract}

The two-step cascade method previously used in neutron capture experiments is now applied to a proton capture
reaction. The spectrum of two-step cascades populating the first $2^+$ level of $\rm ^{60}Ni$ has been
measured with $\rm ^{59}Co(p,2\gamma)^{60}Ni$ reaction. The simulation technique used for the spectrum
analysis allows one to reveal the range of possible shapes of both $E1$ and $M1$ $\gamma$-strength functions.
The low-energy enhancement previously observed in $\rm ^3He$ induced reactions  is seen to appear in $M1$
strength functions of $^{60}$Ni.

\end{abstract}


\maketitle

\section{Introduction}
The $E1$ and $M1$ $\gamma$-strength functions below the particle separation threshold are still a
subject of investigation and a source of large uncertainties in reaction cross-section calculations
\cite{Rauscher}. Despite the long history of experimental studies, no definite results have been
established for this energy region. The current status of $\gamma$-strength functions is summarized
in Ref.~\cite{RIPL}, where the $E1$ strength is described on the basis of a low-energy
extrapolation of the giant dipole resonance (GDR). Usually a modified Lorentz function is applied,
taking into account the energy and temperature dependence of the GDR width. The $M1$ strength is
described by the Lorentz function based on the existence of the spin-flip $M1$ resonance. However,
the parameters of this resonance (peak cross-section, width, and centroid) suffer from large
uncertainties.

Experimental information about $\gamma$-strength functions for $\gamma$-transitions below the particle
separation energy can be obtained from measuring the $\gamma$-spectra of different nuclear reactions.
However, since the spectra depend not only on the $\gamma$-strength functions but also on the density of
levels populated by the $\gamma$-transitions, the interpretation of such spectra is difficult. All
$\gamma$-strength function models recommended in Ref.~\cite{RIPL} are in fact dependent on the level density
model applied, since they were basically obtained from $\gamma$-spectra of neutron capture reactions. Thus,
it was necessary to assume a model for the level density below the neutron binding energy in order to derive
a model for the $\gamma$-strength.

At present, there is an increasing interest in nuclear $\gamma$-strength functions. For example, the recent
finding of an $E1$ pygmy dipole resonance below the neutron threshold for $\rm ^{136}Xe$ from $(\gamma \gamma
')$ experiments \cite{savran} and for $\rm ^{117}$Sn \cite{sn117} from $(^3$He,$^3$He$^{\prime}$) experiments
is very intriguing and could have a large impact on reaction rates relevant for nuclear astrophysics. Also,
new experimental techniques which allows one to study $\gamma$-strength functions below the neutron
separation energy with photon absorption reactions are developed \cite{Rusev}.

An unexpected enhancement of the low-energy part of the $\gamma$-strength function has been reported for
nuclei from the medium-mass region such as $\rm^{56,57}Fe$ \cite{Voinov,Emel} and $\rm ^{50,51}V$ \cite{V},
and for the heavier $\rm ^{93-98}Mo$ \cite{Mo}. These results are obtained with $\rm ^3$He-induced reactions
where the particle-$\gamma$ coincidence data has been analyzed with the so-called Oslo method (Oslo-type
experiments). The question of whether this behavior is unique to these nuclei or if it is a general feature
of a certain mass region is open and requires further investigations.


At this point it is clear that there is no single experiment which would give a complete
picture of the $\gamma$-strength functions in a wide energy range. Different experimental techniques
need to be combined to gradually uncover the various structures and the underlying physics of the
$\gamma$-strength function.

One of the experimental techniques successfully applied to test $\gamma$-strength function models is
the method of two-step $\gamma$-cascades (TSC) following thermal neutron capture. It was first
proposed by Hoogenboom \cite{Hoogen} and later developed by the Dubna group \cite{Bon} and the Prague group
\cite{Becvar}. The idea of this method is to obtain spectra of cascade $\gamma$-transitions
proceeding between compound levels and one of the low-lying discrete levels of the final nucleus.
The sum energy of such two-step cascades is fixed and the shape of the TSC spectra is determined by
the level density and the $\gamma$-strength function of the final nucleus. This method has proven to
be a relatively sensitive test. However, it might suffer from large uncertainties mainly due to unknown level
density functions (one must usually rely on models) and large Porter-Thomas (PT) fluctuations
of the cascade intensities. Therefore, the description of the shape of the TSC spectra can be rather uncertain.
For the absolute normalization of the
TSC intensities one uses literature data on absolute intensities of primary $\gamma$-transitions and
branching ratios of low-lying levels. Both these quantities, especially the latter, are usually not
known with high enough precision.

In order to reduce the uncertainties related to the TSC method for neutron capture reactions, we
measured the TSC spectrum from the proton capture reaction $^{59}$Co$(p,2\gamma)^{60}$Ni. In this work, the
novel approach is that the level density of the final nucleus $\rm^{60}Ni$ has been obtained
by us in an independent experiment utilizing neutron evaporation spectra. This simplified the
analysis of the TSC spectrum considerably. Also, the great advantage of proton capture is that due to
the proton energy spread in the target, many compound nuclear resonances are excited. Thus, a
good averaging of the resonance decay properties is provided and the PT fluctuations are reduced.

In this article we describe our first results on TSC $\gamma$-transitions populating the first
$\rm 2^+ $ level of the final $\rm ^{60}Ni$ nucleus from the proton capture on $\rm ^{59}Co$.

\section{Experiment and method}

Cascade $\gamma$-transitions following the proton capture on $\rm ^{59}Co$ have been measured with
two high-purity germanium (HPGe) detectors (40\% and 60\% efficiency) placed at about 8~cm from the target, making an angle of
about $\rm 125^{\circ}$ to reduce possible anisotropy effects of the cascades caused by
angular correlations. Lead discs of 1-mm thickness  were placed in front of each detector in order to reduce
cross-talk effects. The tandem accelerator of Edwards Accelerator Laboratory, Ohio University, delivered
the proton beam with energy of 1.85 MeV, which is just below the $(p,n)$
reaction threshold. At this energy, along with the $(p,\gamma)$ channel, only the $
(p,\alpha)$ channel is open. The latter channel does not produce significant $\gamma$ emission because of its
small cross section. A 1-$\mu$m thick natural cobalt foil (100\% $^{59}$Co) was used as
target. The energy loss in the target was estimated to be around 80~keV. The beam current was
maintained at about 150-200 nA to keep the counting rate of each detector at about 3000/sec. Single
spectra and coincident events were accumulated simultaneously. Considering the energy spread due
to the energy loss in the target, the number of proton-capture resonances excited are
estimated to be $\rm \sim~70$. Therefore, a good averaging is provided, which ensures
a sufficient independence of the decay properties of the individual resonances.

\begin{figure}
\includegraphics[width=9cm]{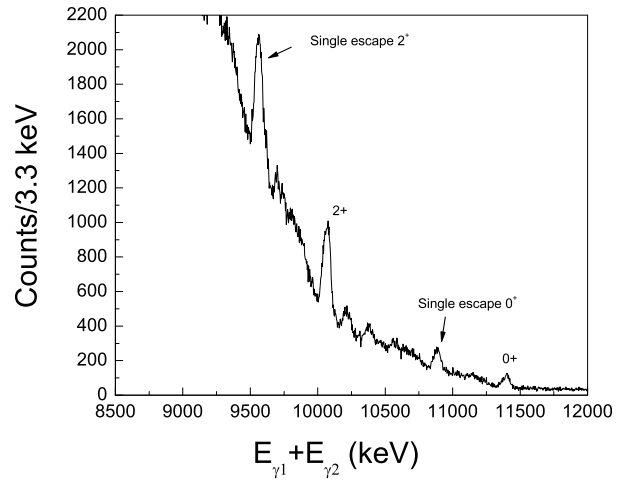}
\caption{The spectrum of the sum amplitude of two coincident pulses from the $
^{59}$Co$(p,2\gamma)^{60}$Ni reaction.} \label{fig:fig1}
\end{figure}

\begin{figure}
\includegraphics[width=9cm]{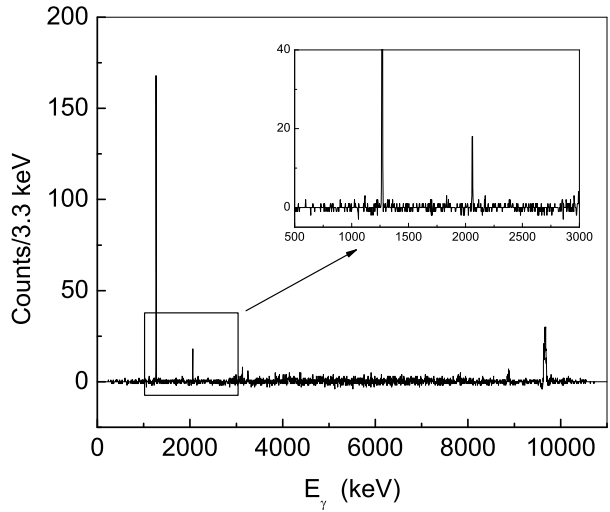}
\caption{The spectrum of two-step $\gamma$-transitions populating the $\rm 2^+$ level of the $\rm
^{60}Ni$ nucleus.} \label{fig:fig2}
\end{figure}

The spectrum of the sum amplitude of two coincident pulses is shown in Fig.~\ref{fig:fig1}.
Two-step $\gamma$-cascades create peaks in such a spectrum revealing the population of discrete
low-lying levels. One can see a very small peak at 11.3~MeV corresponding to the population of the
ground state of $\rm ^{60}Ni$, the large peak populating the first $\rm 2^+$ excited state as
well as satellite peaks caused by single-escape annihilation quanta. These peaks have a width of
around 80 keV, which is consistent with the proton energy loss in the target. This width
is considerably larger than for similar experiments with thermal neutrons, where the peak widths are limited by
the detector resolution only. The Compton background increases considerably as the
energy decreases. Thus, only the $\rm 2^+$ peak has a reasonable peak-to-background ratio in this
experiment. This peak was therefore used to create the corresponding TSC spectrum. This
spectrum is made by placing a gate on the peak and subtracting the spectra obtained by gating on
the background at both sides of the peak (for details, see~\cite{Becvar}). The resulting cascade spectrum
corrected for the detector efficiencies is shown in Fig.~\ref{fig:fig2}.
This spectrum consists of the full-energy absorption peaks only, which are well separated in the
low-energy range and become a smooth distribution towards the center of the spectrum. The low-energy part
of the spectrum is free from any cross-talk effects including low-energy bremstrahlung
as discussed in Ref.~\cite{brem}.

The efficiency of the detectors has been determined by measuring the $\gamma$-spectrum from the
$^{27}$Al$(p,\gamma)$ reaction at $\rm E_p=0.992~MeV$ and using the  standard $\gamma$-intensities for this
reaction from Ref.~\cite{pgst}.

 Ideally, the TSC spectrum should be symmetric in respect to half of the sum energy of two cascades $
(E_{\gamma1}+E_{\gamma2})/2$ because each cascade creates full absorption peaks at two places in the
spectrum, at $ E_{\gamma1}$ and at $ E_{\gamma2}$. However, due to different energy resolutions, the
high-energy discrete peaks are considerably broader compared to their low-energy satellites.

We also accumulated the singles $\gamma$-ray spectrum with energies from ~0.3 to 12~MeV. The primary interest
here is the $2^+ \rightarrow 0^+$ 1332-keV $\gamma$ transition since its absolute intensity consists of more
than 90\% of the compound $\rm ^{60}Ni$ nucleus decays. The intensity of cascade transitions was determined
relative to the intensity of this 1332~keV $\gamma$-transition. In the following, this normalized TSC
spectrum is analyzed to obtain the $\gamma$-ray strength function.

\section{Comparison with model calculations}

In order to reduce the PT fluctuations of the individual cascades, the TSC spectrum in Fig.~\ref{fig:fig2}
was compressed into 500-keV energy bins. Assuming that the proton capture at
this energy is due to the compound reaction mechanism, the shape and the absolute intensity of the
 spectrum is determined by the $E1$, $M1$, and (to a lesser extent) the $E2$ strength functions of both
the primary and the secondary $\gamma$-transitions in the cascades. It is also determined by the level
density of the residual nucleus $\rm ^{60}Ni$.

\subsection{Level density and $\gamma$-strength functions}
Traditionally, the level density was one of the most uncertain quantities when analyzing
$\gamma$-spectra from nuclear reactions. In our case the level density has been obtained by us from
the reactions $^{59}$Co$(d,n)^{60}$Ni, $^{58}$Fe$(^3$He$,n)^{60}$Ni \cite{Voinov} and $^{55}$Mn$(^6$Li$,n)^{60}$Ni
\cite{CT}. Based on these experiments, it turned out that it is more appropriate to use
the constant-temperature level density
$\rm \rho(U)=1/T \exp((E-E_0)/T)$ (compared to the Fermi-gas model with $T \sim \sqrt E$)
with a temperature of $\rm T=1.4$~MeV and
an energy shift of $\rm E_0=-0.85~MeV$.

\begin{figure}[t]
\centering
\includegraphics[width=9cm]{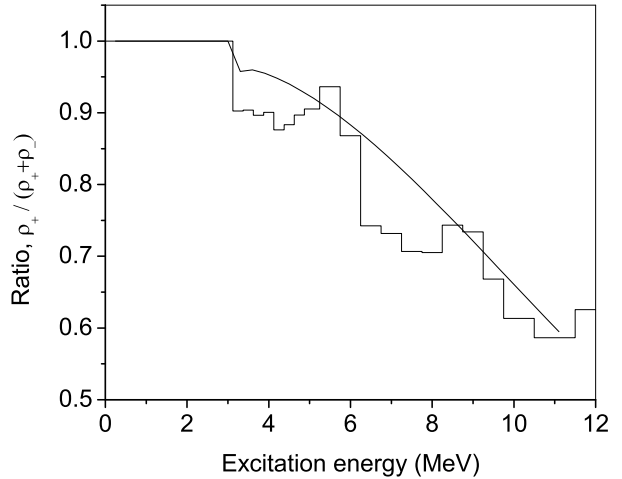}
\caption{Ratio of the density of positive parity levels to the total level density for $\rm ^{60}Ni$.
The histogram represents the calculations of Ref.~\cite{Goriely1}. The line shows the approximation used in our
calculations.} \label{fig:fig5}
\end{figure}

Another important feature affecting the calculations is the ratio of levels with positive and
negative parities. Usually, the number of positive and negative levels are assumed to be equal in
level-density model calculations. However, this is not the case for the $\rm ^{60}Ni$ where, as
seen from the level scheme, the deviation from the parity balance is obvious and can no longer be
neglected. Up to 4.2~MeV of excitation energy, $\rm ^{60}Ni$ has only one level of negative parity,
while the rest have positive parity. The parity ratio above 4.2~MeV must be estimated on the basis
of model calculations. We have used the calculations of Ref.~\cite{Goriely1} based on the HF-BCS
microscopical approach. The result of these calculations for $\rm ^{60}Ni$ and the approximation we
have applied in our TSC calculations are presented in Fig.~\ref{fig:fig5}.

We have tested the most commonly used prescriptions for the $\gamma$-strength functions. The first and most
traditional one is the standard Lorentzian (SLO) function  for $E1$ transitions with parameters
fitted to the GDR \cite{RIPL}.  There are also models which take into account
the temperature dependence of the $\gamma$-strength such as the Kadmensky-Markushev-Furman (KMF) model
\cite{Kad83} and the generalized Lorentzian (GLO) \cite{Kopecky}. The $M1$ and $E2$
$\gamma$-strength functions are more uncertain. For both of them we used the standard Lorentzian
function with parameters based on the recommended systematics of Ref.~\cite{RIPL}. For the $M1$
strength, the single-particle (SP) energy independent function \cite{blatt} has been tested as well.

The spin distribution of the compound levels populated through 1.85-MeV proton capture can be estimated on the
basis of a proton optical potential which determines the transmission coefficients for each orbital
momentum of the captured proton. There are three different potentials available in the RIPL data
base \cite{RIPL}. For these three potentials, we estimated the following fractions of the capture
cross section for the different $\ell$ values of the captured protons: 0.64, 0.55 and 0.52 for s-wave protons;
0.15, 0.27 and 0.30 for p-wave protons; and 0.20, 0.17, and 0.17 for d-wave protons. We tested all three
potentials in our simulations and found that, within the uncertainties, our final results did not depend on
the particular potential used. The results we present here is based on the optical potential of Ref.~\cite{Kon}
(corresponding to the first numbers of the given fractions above).

%
\begin{table}
\caption{Ratio of calculated and experimental intensities for cascades populating the $\rm 2^+$ level
of $\rm ^{60}Ni$. }
\begin{tabular}{|c|c|l|c|} \hline
\multirow{2}{*}{$E1$+$M1$ models}        &\multicolumn{3}{|c|}{Energy interval,MeV}  \\ \cline{2-4}
   &0.5-2 &2.15-3.4 &4-5.1 \\ \hline
  KMF+SLO      & 0.87(17)  &3.10(120)         &1.84(37)      \\
  KMF+SP       & 0.58(12)  &1.74(70)         &1.08(21)     \\
  SLO+SLO      & 0.64(13)  &2.45(100)         &1.58(32)      \\
  SLO+SP       & 0.48(10)  &1.56(60)         &1.10(22)      \\
  GLO+SLO      & 0.51(10)  &1.81(72)         &1.24(25)      \\
  GLO+SP       & 0.38(8)  &1.11(44)        &0.71(14)      \\ \hline

\end{tabular}
\label{tab:parameters}
\end{table}

In order to compare the experimental TSC spectrum with the calculations we have chosen three energy
intervals of the spectrum of Fig.~\ref{fig:fig2}. These intervals reflect the most important
features of the spectrum and determine the spectral shape. One can see from Table~\ref{tab:parameters}
that the main
problem of all the models is that they are not able to simultaneously describe the first and the
second energy interval of the spectrum, although many models describe well the cascade intensity in
the third energy interval belonging to the middle part of the spectrum. This result shows that
analyzing just the middle part of the spectrum is not sufficient to unambiguously test
$\gamma$-strength functions with TSC spectra. Here we can use the advantage of the $
(p,2\gamma)$ reaction compared to the $ (n,2\gamma)$ reaction: in the case of neutron capture
reactions only the middle part of the TSC spectra is usually compared with calculations due to
increasing PT fluctuations in other energy intervals \cite{Becvar}.

\subsection{Simulations}

In order to find the range of possible $E1$ and $M1$ $\gamma$-strength function shapes which would describe
the TSC spectrum, we simulated the TSC spectrum with randomly shaped input strength functions. The range of
input strength function shapes has been determined taking into account the current knowledge about the
possible forms of their energy dependence. Since most models are based on the Lorentz function, we have
chosen this as a basic function to describe both $E1$ and $M1$ $\gamma$-strength below the particle
threshold. In addition to that, an exponential low-energy enhancement function has been added to mimic the
possible alteration of the Lorentz function in the low-energy intervals. The possible uncertainties in the
general slope have been simulated using the multiplication term of the form $\exp[A (E_\gamma-11.3)]$, where
the number $\rm 11.3$ is chosen to be equal to the excitation energy (in MeV) of the compound nucleus and $A$
is the coefficient determining the slope. This gives the functional form of both $E1$ and $M1$ strength
functions as:

\begin{eqnarray}
 f=8.68\cdot10^{-8}{(\rm mb^{-1}~MeV^{-2})} \times \nonumber \\
 C_{M1} [\frac{E_\gamma \Gamma}{(E_\gamma^2-E_r^2)^2+E_\gamma^2\Gamma}+
 B  \exp(-CE_\gamma)]\times \nonumber \\
 \exp\left[ A(E_\gamma-11.3) \right]
 \label{eq:ran}
\end{eqnarray}

\begin{figure*}[t]
\centering
\includegraphics[width=16cm]{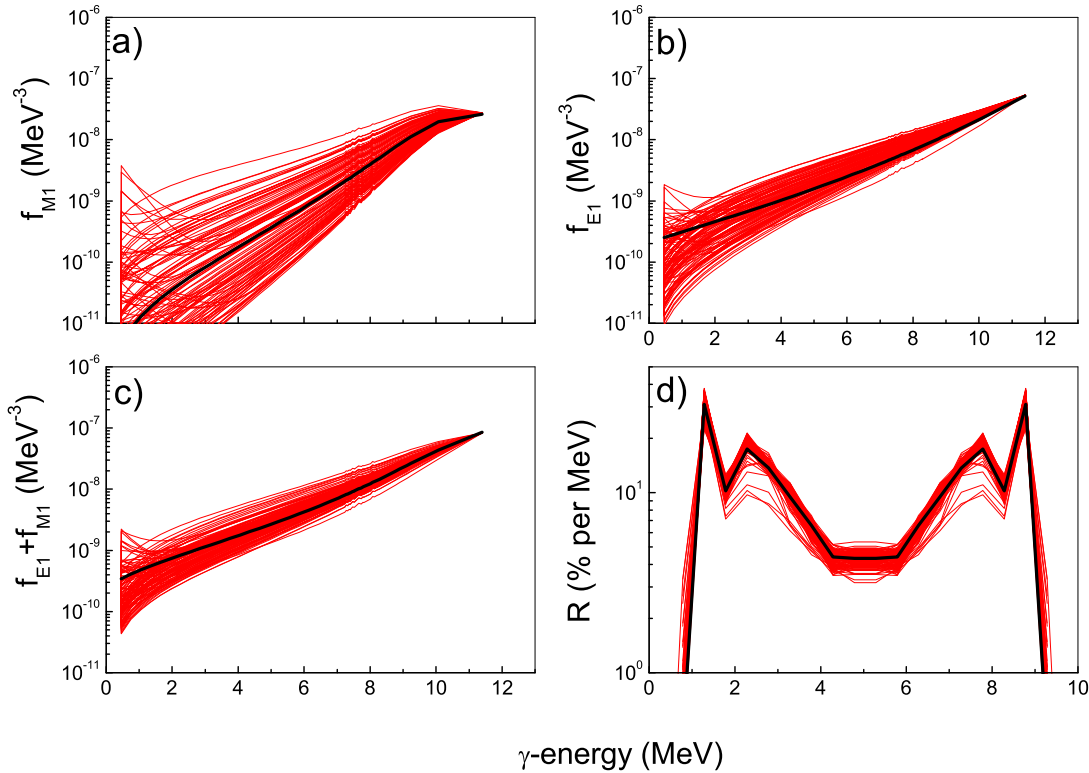}
\caption{(Color online) Results of the simulations of the two-step cascade spectrum populating the $2^+$
level of $\rm ^{60}Ni$. The black thick lines in the panels a), b), and c) are input strength functions which
result in the black TSC spectrum in panel d). The thin red lines show the output functions and the
corresponding TSC spectra resulting from the simulation procedure. In panel d), $R$ is the ratio of the TSC
intensity and the intensity of the $\rm 2^+ \rightarrow 0^+$ ground state transition in $\rm ^{60}Ni$. Note
that the absolute scale of the strength functions is approximate.} \label{fig:fig3}
\end{figure*}
The parameters $A$, $B$, and $C$ were varied randomly to accommodate a wide range of function shapes. The
parameter $C_{M1}$ (applied to the $M1$ strength function) varied randomly between 0.1 and 0.7 to be
consistent with the experimental systematics and the corresponding uncertainties of the $f_{M1}/f_{E1}$ ratio
\cite{RIPL}. The parameters for the GDR has been taken from Ref.~\cite{RIPL}. It has been tested that the
function given by Eq. (\ref{eq:ran}) is able to mimic all known functions currently used to model
$\gamma$-strength functions, including those used previously in this paper (see Table~\ref{tab:parameters}).
The only restriction is that it assumes that the Axel-Brink hypothesis is valid, which means that the
strength functions do not depend on the excitation energy of the final levels populated by the
$\gamma$-transitions. This is not the case for the KMF and GLO models in which the strength depends on the
temperature, which in turn depends on the intrinsic excitation energy $U$ as the Fermi-gas model predicts
($T=\sqrt{(U/a)}$, where $a$ is the level-density parameter). The temperature behavior in nuclei is an open
problem and there is no conclusive experimental data in the energy range of our interest. For example,
uncertainties regarding the Fermi-gas versus the constant-temperature level density models still exist. In
the case of $\rm ^{60}Ni$, we have showed in a previous work that the constant-temperature model works better
for excitation energies up to 20~MeV \cite{CT}. Therefore we deem that the Axel-Brink hypothesis is justified
for this specific nucleus in the energy region of interest.

We have simulated $E1$ and $M1$ strength functions according to Eq.~(\ref{eq:ran}). The $E2$ strength
function has been calculated using a Lorentz function with parameters according to the systematics of
Ref.~\cite{RIPL}. For each combination of random $E1$ and $M1$ strength functions the TSC spectrum is
generated. Only those spectra that agree with the experimental one within predefined uncertainties are
selected. The uncertainties have been estimated to be 20, 40 and 20 percent for the first, second, and third
energy intervals, respectively (see Table~\ref{tab:parameters}).

First, to investigate how well the simulation procedure works, we used a test spectrum calculated with a set
of selected input strength function models to check that the input and output strength functions are
consistent. The models we chose to use were the GLO model with a constant temperature for the $E1$ strength,
and the SLO model for the $M1$ strength function. Then, we simulated this spectrum with random strength
functions using Eq.~(\ref{eq:ran}). The output strength functions that reproduce the test spectrum within the
predefined uncertainties were selected. These strengths are compared with the input strength functions in
Fig.~\ref{fig:fig3}. The absolute normalization of the strength functions does not affect the TSC
calculations, only the slope in important. Therefore, in order to reveal the shape variation, all strength
functions have been normalized to the same number at $\rm E_\gamma \approx 11.3~MeV$, which is approximately
consistent with the absolute scale at these $\gamma$-energies.
\begin{figure*}[t]
\begin{center}
\includegraphics[width=16cm]{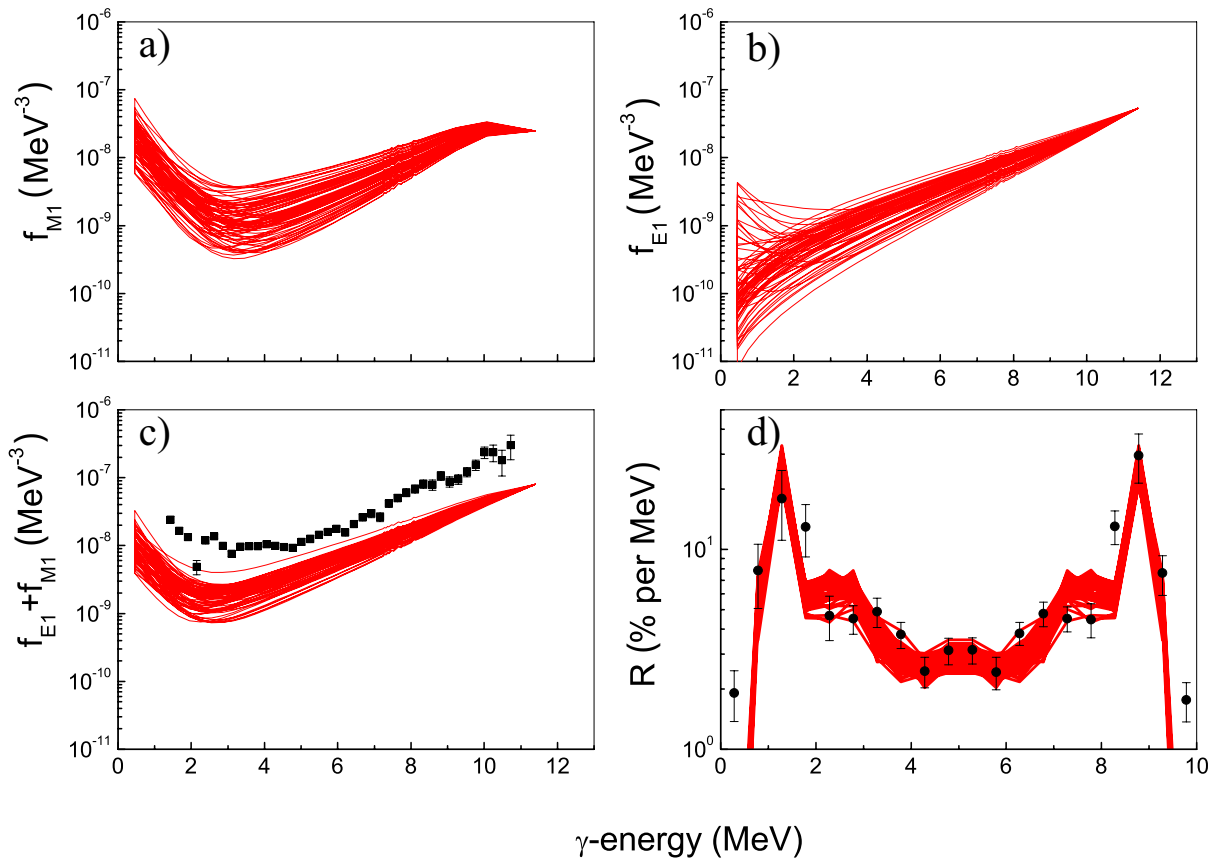}
\caption{Same as in Fig.~\ref{fig:fig3} but for the experimental TSC spectrum. The black points in
panel c) are data on $^{56}$Fe from the Oslo experiment \cite{Emel}.} \label{fig:fig4}
\end{center}
\end{figure*}
One can see the degree of sensitivity of the cascade spectrum to the functional dependence of the strength
functions. Within the adopted uncertainties of the TSC spectrum, the slopes of both the $E1$ and $M1$
strength functions exhibit large variations. However, the important result from this simulation test is that
the original input strength functions are in the uncertainty corridor of the output functions. The spectrum
does not appear to be sensitive to the $\gamma$-strength functions for $\gamma$-ray energies below 2~MeV.
However, one should mention that in general, the degree of sensitivity can vary for different
nuclei.

Next, we performed a simulation in the same way as previously described, but now we selected only those
output strength functions which reproduce the experimental TSC spectrum within our predefined uncertainties
in the three energy intervals. The result is shown in Fig.~\ref{fig:fig4}. Although there are large
uncertainties  in the obtained functions, it is obvious that all the possible $M1$ strength functions show a
low-energy increase. However, the $E1$ function below $\rm \sim 2~MeV$ is rather uncertain; this strength
might have an increase at low energies, but it is equally probable to have a decreasing $E1$ strength. The
sum of $E1$ and $M1$ functions exhibits less uncertainties and therefore the low-energy increase is more
pronounced here. This shape is consistent with what we have observed previously in Oslo-type experiments for
nuclei in this mass region, $\rm ^{56}Fe$ and $\rm ^{57}Fe$ \cite{Voinov,Emel}. However, from the current
simulation we can specify that the low-energy enhancement is most likely due to $M1$ $\gamma-$transitions.

\section{Discussion}

The performed simulations show that the proton-capture TSC spectra can be equally well described with a
variety of $E1$ and $M1$ strength functions. However, some general trends of the $\gamma$-strength functions
can be studied unambiguously with this technique.

Because there are many resonances excited in a proton capture experiment, the obtained TSC spectra are less
vulnerable to PT fluctuations compared to similar spectra from thermal neutron-capture experiments. This fact
considerably reduces the uncertainties of the resulting $\gamma$-strength functions. Indeed, we have
performed additional simulations where only the third energy interval corresponding to the middle of the TSC
spectrum was used to select output strength functions. The middle interval is traditionally used to compare
experimental and theoretical TSC spectra obtained from thermal neutron capture experiments \cite{Becvar},
since in this part of the spectrum, the number of intermediate levels increases to the point where
Porter-Thomas fluctuations are sufficiently reduced so that they become comparable to experimental errors.
However, if only this energy region is used, the uncertainties of the shapes of the output strength functions
become considerably larger.

Detailed calculations show that the main contribution to the TSC intensity comes from cascades with hard
primary transitions whose energies exceed half of the sum of the cascade energies, i.e.
$E_{\gamma1}>(E_{\gamma1}+E_{\gamma2})/2$. These transitions populate the lower half of the accessible
excitation-energy region of $\rm ^{60}Ni$, where levels of positive parity dominate. Taking into account that
for this reaction, s-wave protons populate $\rm 3^-$ and $4^-$ spins in about 90\% of the cases, the cascades
populating the $2^+$ final level have $E1$+$M1$ multipolarity with $ E_{\gamma2}^{M1}<~ 5~MeV$. That is why
the TSC spectrum in our experiment is sensitive to low-energy $M1$ strength. In other nuclei the situation
might be different.

\section{Conclusion}

In this paper the two-step cascade method previously developed and used for neutron-capture experiments, has
been applied to proton-capture reactions. The advantage of the proton  TSC spectra is the possibility to get
the absolute TSC intensities with much greater precision compared to TSC intensities from neutron-capture
reactions. Secondly, the proton TSC spectra undergo less PT fluctuations. This fact allows us to put more
restrictions on the range of possible $\gamma$-strength functions describing the experimental TSC spectra.

A simulation technique has been used to infer strength functions capable to describe the experimental
spectrum. The obtained $M1$ strength functions  support strongly a low-energy increase resulting in a similar
increase in the sum of the $E1$+$M1$ strength functions. This increase is consistent with what is observed
for nuclei in the same mass region (iron nuclei from Oslo-type experiments \cite{Voinov,Emel}). However, the
TSC spectrum from the $ ^{59}$Co$(p,2\gamma)$ reaction is not sensitive enough to determine whether there is
a low-energy enhancement of the $E1$ $\gamma$-strength function.

In conclusion, we have demonstrated in this work that the combination of analyzing TSC spectra from
proton-capture reactions along with measuring level densities from particle-evaporation experiments is a good
supplementary tool to study both $E1$ and $M1$ strength functions in atomic nuclei.


\end{document}